\documentclass[]{pasj01}
\usepackage{color}

\begin{document} 
\Received{}
\Accepted{}

\title{The nature of H$\alpha$-selected galaxies along the large-scale structure at z=0.4 revealed by Subaru Hyper Suprime-Cam survey}


\author{Yusei \textsc{Koyama}\altaffilmark{1,2}}%
\altaffiltext{1}{Subaru Telescope, National Astronomical Observatory of Japan, National Institutes of Natural Sciences, 650 North A'ohoku Place, Hilo, HI 96720, U.S.A.}
\altaffiltext{2}{Graduate University for Advanced Studies (SOKENDAI), Osawa 2-21-1, Mitaka, Tokyo 181-8588, Japan}
\email{koyama@naoj.org}

\author{Masao \textsc{Hayashi}\altaffilmark{3}}
\altaffiltext{3}{National Astronomical Observatory of Japan, Osawa 2-21-1, Mitaka, Tokyo 181-8588, Japan}

\author{Masayuki \textsc{Tanaka}\altaffilmark{3}}

\author{Tadayuki \textsc{Kodama}\altaffilmark{3,4}}
\altaffiltext{4}{Astronomical Institute, Tohoku University, 63 Aramaki, Aoba-ku, Sendai 980-8578, Japan}

\author{Rhythm \textsc{Shimakawa}\altaffilmark{5}}
\altaffiltext{5}{UCO/Lick Observatory, University of California, 1156 High Street, Santa Cruz, CA 95064, U.S.A.}

\author{Moegi \textsc{Yamamoto}\altaffilmark{2}}

\author{Fumiaki \textsc{Nakata}\altaffilmark{1}}

\author{Ichi \textsc{Tanaka}\altaffilmark{1}}

\author{Tomoko \textsc{Suzuki}\altaffilmark{3}}

\author{Ken-ichi \textsc{Tadaki}\altaffilmark{3}}

\author{Atsushi J. \textsc{Nishizawa}\altaffilmark{6}}
\altaffiltext{6}{Institute for Advanced Research, Nagoya University, Nagoya 464-8602, Aichi, Japan} 

\author{Kiyoto \textsc{Yabe}\altaffilmark{7}}
\altaffiltext{7}{Kavli Institute for the Physics and Mathematics of the Universe, The University of Tokyo, 5-1-5 Kashiwanoha, Kashiwa, Chiba 277-8583, Japan}

\author{Yoshiki \textsc{Toba}\altaffilmark{8}}
\altaffiltext{8}{Academia Sinica Institute of Astronomy and Astrophysics, P.O. Box 23-141, Taipei 10617, Taiwan}

\author{Lihwai \textsc{Lin}\altaffilmark{7}}

\author{Hung-Yu \textsc{Jian}\altaffilmark{7}}

\author{Yutaka \textsc{Komiyama}\altaffilmark{3,2}}



\maketitle

\begin{abstract}
We present the environmental dependence of colour, stellar mass, and star formation (SF) activity in H$\alpha$-selected galaxies along the large-scale structure at z=0.4 hosting twin clusters in DEEP2-3 field, discovered by Subaru Strategic Programme of Hyper Suprime-Cam (HSC SSP). By combining photo-$z$ selected galaxies and H$\alpha$ emitters selected with broad-band and narrow-band (NB) data of the recent data release of HSC SSP (DR1), we confirm that galaxies in higher-density environments or galaxies in the cluster central regions show redder colours. We find that there still remains a possible colour--density and colour--radius correlation even if we restrict the sample to H$\alpha$-selected galaxies, likely due to the presence of massive H$\alpha$ emitters in denser regions. We also find a hint of increased star formation rates (SFR) amongst H$\alpha$ emitters towards the highest-density environment, again primarily driven by the excess of red/massive H$\alpha$ emitters in high-density environment, while their specific SFR does not significantly change with environment. This work demonstrates the power of the HSC SSP NB data to study SF galaxies across environment in the distant universe. 
\end{abstract}

\section{Introduction}

In the $\Lambda$CDM universe, small structures form first, and then they merge to form larger scale systems. Clusters of galaxies are therefore the most massive systems in the universe, while clusters also grow into larger-scale structures (or super clusters) by merging themselves (\cite{springel2005}). It is well established that galaxy properties significantly change along the cosmic web, from rich clusters/groups to general fields. The well-known morphology--density  (e.g. \cite{dressler1980}; \cite{goto2003}; \cite{bamford2009}), colour--density (e.g.\ \cite{tanaka2004}; \cite{baldry2006}; \cite{haines2007}) or star formation rate (SFR)-density (e.g.\  \cite{lewis2002}; \cite{gomez2003}) correlations seen in the local universe are expected to be established during the course of cluster scale assembly. However, the key physical mechanisms responsible for the environmental effects (or environmental quenching) are still unclear.  

Observations of large-scale structures in the distant universe provide us with a number of insights into how galaxies grow along the history of the formation of galaxy clusters or groups. The Subaru Prime Focus Camera (Suprime-Cam; \cite{miyazaki2002}) played an important role in identifying large-scale filamentary structures in high-$z$ universe, taking advantage of its wide field of view and large light collecting power of the Subaru Telescope. \citet{kodama2001}, \citet{tanaka2005}, and \citet{koyama2008} studied colour--density relation for galaxies around rich X-ray selected clusters at $z$$=$0.4--0.8 with Suprime-Cam, and all of these studies demonstrate that galaxy colours start changing at relatively low-density groups or filaments, suggesting that galaxy transition takes place not only in extremely high-density environments like X-ray luminous cluster cores, but also in relatively poor group-scale environments. 

Narrow-band (NB) imaging, in particular combined with a wide-field camera such as Suprime-Cam, is an efficient and powerful approach to identify star-forming (SF) galaxies along the cosmic web in the distant universe based on their strong emission lines (e.g.\ \cite{shimasaku2003}; \cite{kodama2004}; \cite{hayashino2004}; \cite{matsuda2004}; \cite{ouchi2005}; \cite{ideue2009}; \cite{hayashi2010}; \cite{koyama2011};  \cite{sobral2011}; \cite{matsuda2011}; \cite{tadaki2012}; \cite{yamada2012}; \cite{toshikawa2012}; \cite{kajisawa2013}; \cite{darvish2014}; \cite{shimakawa2017}; \cite{stroe2017}). Unlike multi-object spectroscopy, NB imaging technique allows one to construct a complete sample of galaxies down to a certain level of emission line flux (hence SFR) over the observed field of view. Because emission line fluxes, most preferably H$\alpha$$\lambda$6563 emission line, can directly be translated into SFRs, we can study galaxies along the large-scale structures in the distant universe in a more quantitative way than the simple broad-band colour approaches which suffer from degeneracy between age, metallicity, and dust extinction. It is true that even H$\alpha$ line could significantly underestimate SFRs due to dust extinction for extremely dusty and/or highly star-bursting population (e.g.\ \cite{puglisi2017}), but the NB H$\alpha$ survey is still the most efficient way to study SF galaxies within a narrow redshift slice.

\citet{koyama2011} performed a wide-field NB H$\alpha$ imaging survey of a very rich cluster at $z=0.4$ (Abell~851) using Suprime-Cam, and find that star-forming galaxies showing red colours are strongly clustered in group-scale environment in the cluster outskirts.  Our follow-up study using Spitzer mid-infrared data showed that those red star-forming galaxies are dusty red galaxies, as many of them are individually detected at 24$\mu$m (\cite{koyama2013}). They also showed evidence that H$\alpha$-selected galaxies in $z=0.4$ cluster environments are more highly obscured by dust than field counterparts at fixed stellar mass, by comparing their average IR-based and H$\alpha$-based SFRs. \citet{sobral2016} also independently confirmed this trend using Balmer decrement analysis (H$\beta$/H$\alpha$ ratio) with optical spectroscopic observations of the same targets. 

Star-forming galaxies in general exhibit a tight correlation between SFR and stellar mass ($M_{\star}$); so-called SF main sequence. The location of the SF main sequence changes with cosmic time (e.g. \cite{daddi2007}; \cite{elbaz2007}; \cite{whitaker2012}; \cite{speagle2014}), while there is very little environmental dependence at fixed redshift (e.g.\ \cite{peng2010}; \cite{mcgee2011}; \cite{koyama2013}; \cite{lin2014};  \cite{koyama2014}; \cite{darvish2016}; \cite{duivenvoorden2016}; \cite{wagner2017}). \citet{koyama2013} compared the SFR--$M_{\star}$ relation for H$\alpha$-selected galaxies in clusters and fields at $z$$=$0.4--2.2, and conclude that the environmental variation of the SF main sequence is always small over the last $\sim$10-Gyrs (with $\sim$0.2-dex at maximum). However, because of the limited sample size of H$\alpha$ galaxies available in distant clusters, their work is based on a comparison between cluster galaxies (defined as those within 2~Mpc from the clusters) and field H$\alpha$ emitters (selected from HiZELS; \cite{sobral2013}). Extending the study with larger emission-line galaxy samples, covering wide environmental range from rich clusters to isolated field environments, is thus an essential step towards understanding the effects of environment more globally, more completely, and in a more unbiased way. 

As the successor of Suprime-Cam, the new Subaru wide-field optical camera, Hyper Suprime-Cam (HSC; Miyazaki et al. 2017), can now provide us with an ideal tool to study galaxy environment in the distant universe. Because large-scale structures at the intermediate-redshift universe (at $z\sim 1$) are known to be extended over a few degrees scale on sky (e.g.\ \cite{nakata2005}; \cite{tanaka2009}), the unprecedentedly wide field coverage (1.5~deg) of HSC allows one to efficiently study galaxies across wide environmental range.  In this paper, we present an initial result on the environmental dependence of the properties of star-forming galaxies based on H$\alpha$-selected galaxies at $z=0.4$ selected with broad-band and narrow-band data of the recent data release of HSC Subaru Strategic Programme (SSP) (\cite{aihara2017a}; \cite{aihara2017b}). The goal of this paper is to study colour, stellar mass, and star-forming activity in galaxies over the huge cosmic web at $z=0.4$, by taking full advantage of the new HSC data.  

This paper is organized as follows. In Section~\ref{sec:data}, we briefly describe the HSC SSP survey data used in this study, and summarize our $z=0.4$ H$\alpha$ emitter sample in the DEEEP2-3 field. Our main results are presented in Section~\ref{sec:result}. In Section~3.1 and 3.2, we present a newly discovered super structure at $z=0.4$ in DEEP2-3 field hosting twin clusters revealed by the HSC-SSP survey, and define the environment based on the local galaxy number density and the distance to the twin clusters. After presenting the colour--density and colour--radius relation for all $z\sim 0.4$ galaxies in Section~3.3, we discuss in Section~3.4 and Section~3.5 the environmental dependence of various properties amongst H$\alpha$-selected galaxies along the $z=0.4$ cosmic web. In Section~3.6, we compare our results with previous studies focusing on the environmental effects on star-forming galaxies. In Section~3.7, we discuss the effects of more local environments by studying the properties of H$\alpha$ galaxies as a function of the distance to their nearest neighbour galaxy. Finally, we describe our conclusions in Section~\ref{sec:conclusions}. Throughout the paper, we adopt the cosmological parameters of $\Omega_{\rm{M}} =0.3$, $\Omega_{\Lambda} =0.7$, and $H_0 =70$ km s$^{-1}$Mpc$^{-1}$, which gives a 1$''$ scale of 5.4~kpc at $z=0.4$. All magnitudes are given in the AB system, and we assume the Salpeter IMF \citep{salpeter1955} throughout the paper. 


\section{Data}
\label{sec:data}

\subsection{Selection of H$\alpha$ emitters in DEEP2-3 field from HSC-SSP narrow-band imaging data}

We use the broad-band ($grizy$) and narrow-band (NB921) photometric catalogue from the internal data release of the HSC SSP (S15B) released in January 2016 \citep{aihara2017a}. \citet{hayashi2017} used this dataset to identify narrow-band excess galaxies over 16.2~deg$^2$ in all HSC Ultra-Deep/Deep fields. We refer to \citet{hayashi2017} for details of the procedure of emission-line galaxy selection, but we here briefly summarize our methodology to select H$\alpha$ emitters at $z=0.4$ from NB921 data in the DEEP2-3 field covering 5.6~deg$^2$ around (R.A.=352$^{\circ}$.2, Dec.=$-$0$^{\circ}$.2). This field is one of the HSC-Deep survey layers, where we find a significant over density of H$\alpha$ emitters at $z\sim 0.4$ (Section~3.1). We note that the central wavelength of the NB921 filter is 9214\AA\ with its filter width of 135\AA, corresponding to the H$\alpha$ redshift range of $z$$=$0.393--0.414. 

We first note that the HSC pipeline (hscPipe; Bosche et al. 2017) provides various types of photometric magnitudes in the original HSC catalogue. In this paper, following \citet{hayashi2017}, we use {\sc cmodel} magnitudes, computed by fitting the bulge and disk components for individual sources by taking into account the PSF variation (see Bosche et al. 2017 for more details on the HSC data processing). There are two types of {\sc cmodel} magnitudes: "unforced" and "forced" photometry. The former is independent measurements for individual sources at their position in each band, while the latter performs photometry at the fixed position of the sources determined in the {\it i}-band data\footnote{For sources not detected in $i$-band, the pipeline attempts to determine the source position in $r$-band image as the second priority; the priority is in the order of {\it i, r, z, y, g}. Because we apply broad-band magnitude cut (see Sec.~2.1), all galaxies used in this paper are detected at $i$-band.}. Following \citet{hayashi2017}, we use {\sc cmodel} "unforced" magnitudes for selecting NB emitters, while we use {\sc cmodel} "forced" magnitudes when we measure broad-band colours. 

\begin{figure}
 \begin{center}
\vspace{-2mm}
\includegraphics[width=7.5cm,angle=0]{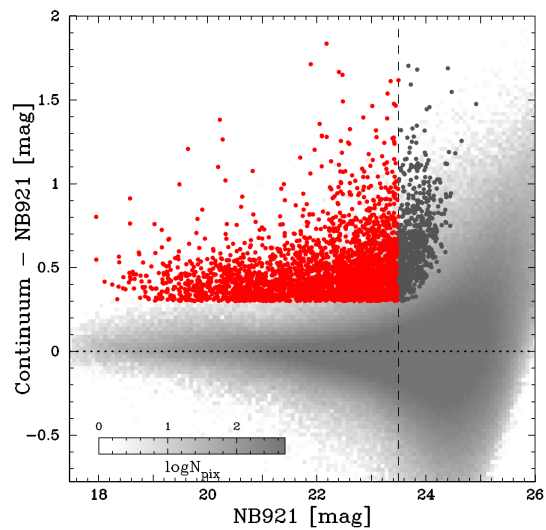} 
 \end{center}
\vspace{-1mm}
\caption{Narrow-band excess (with respect to the continuum magnitude) plotted against the NB921 magnitudes for galaxies in the DEEP2-3 field. The red and dark-grey data points show the H$\alpha$ emitters at $z=0.4$ selected by \citet{hayashi2017} with NB921$<$23.5 and NB921$>$23.5, respectively. We note that galaxies with NB921$<$23.5 are used in this study. The background grey scale shows the distribution of all galaxies in the DEEP2-3 field. To make this grey-scale plot, we apply 120$\times$120 gridding and count the number of data points in each pixel. We follow this strategy in the following plots, by adjusting the grid size according to the number of data points in each plot.}\label{fig:emitter_check}
\end{figure}
\begin{figure*}
 \begin{center}
\vspace{-2mm}
\includegraphics[width=16.0cm,angle=0]{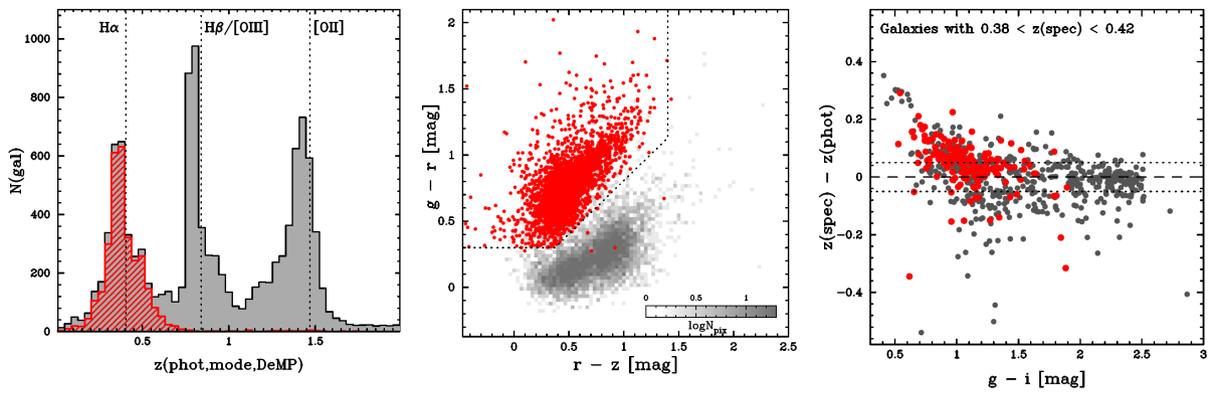}
 \end{center}
\vspace{-1mm}
\caption{ (Left): The distribution of photo-$z$ for all NB921 emitters (grey histogram) and the H$\alpha$ emitters at $z=0.4$ selected by \citet{hayashi2017} (red hatched histogram). 
(Middle): Colour--colour diagram ($g-r$ versus $r-z$) for the H$\alpha$ emitters with NB921$<$23.5 (red circles) and all possible NB921 emitters in the DEEP2-3 field (in grey scale). 
(Right): Difference between $z_{\rm spec}$ and $z_{\rm phot}$ plotted against $g-i$ colours for galaxies with spectroscopic redshift of 0.38$<$$z_{\rm spec}$$<$0.42 in the DEEP2-3 field. The red symbols show the spectroscopically confirmed H$\alpha$ emitters at $z=0.4$ in the same field. The horizontal dotted lines ($\Delta$$z_{\rm phot}$$=\pm$0.05) show our photo-$z$ slice applied to select photometric members in this paper. }\label{fig:photoz_cut}
\end{figure*}

After carefully masking the images and removing junk/stellar objects (including cosmic rays, transient/variable sources; see \cite{hayashi2017}), they first determine the underlying continuum level at the wavelength of NB921 to correct for the colour term to account for the small difference in the effective wavelengths between broad-band ($z$) and narrow-band (NB921) filters. \citet{hayashi2017} use the following equation to estimate the continuum levels at the wavelength of NB921: 
\begin{equation}
m_{\rm NB921, cont} = m_{z} - (0.11 \times (m_{i} - m_{y}) -0.023).
\end{equation}
They define the NB921 emitters as those having $m_{\rm NB921, cont} - m_{\rm NB921}$$>$0.3~mag and significant ($>$5$\sigma$) NB excess with respect to the continuum flux. 

Fig.~\ref{fig:emitter_check} shows narrow-band excess ($m_{\rm NB921, cont} - m_{\rm NB921}$) against the NB921 magnitudes for galaxies in the DEEP2-3 field. To secure the completeness, we apply a magnitude cut of NB921$<$23.5 mag (as shown with the vertical dashed line in Fig.~\ref{fig:emitter_check})\footnote{\citet{hayashi2017} investigate the number counts of NB-detected sources, and determine the completeness limit at the magnitude where the number counts starts to drop (see \cite{hayashi2017} for details).}. Furthermore, we also apply broad-band magnitude cut of $g$$<$26.8, $r$$<$26.6, $i$$<$26.5, $z$$<$25.6, and $y$$<$24.8, which correspond to the 5$\sigma$ limiting magnitude of the HSC-Deep broad-band data. We find that the broad-band data are deep enough, and $>$99\% of the sources with NB921$<$23.5 satisfy the above broad-band criteria. 
 
For the NB921 emitters selected above, \citet{hayashi2017} determine the redshift of each galaxy by investigating their spectroscopic redshifts ($z_{\rm spec}$), photometric redshifts ($z_{\rm phot}$), and broad-band colours. As a sanity check, we show in Fig.~\ref{fig:photoz_cut} (left) the photo-$z$ distribution of the NB921 emitters in the DEEP2-3 field computed by "DEmP" code (\cite{hsieh2014}). It is expected that the majority of the NB921 emitters are H$\alpha$ emitters at $z\sim 0.4$, H$\beta$/[OIII] emitters at $z\sim 0.8$, or [OII] emitters at $z\sim 1.5$, which is recognized as three redshift peaks in the grey histogram in Fig.~\ref{fig:photoz_cut} (left). For the NB921 emitters whose photo-$z$s are not consistent with $z=0.4$, we investigate their broad-band colours to "rescue" possible H$\alpha$ emitters. Fig.~\ref{fig:photoz_cut} (middle) shows $g-r$ versus $r-z$ diagram for all NB921 emitters in the DEEP2-3 field, where we show the colour criteria to select H$\alpha$ emitters defined by \citet{hayashi2017} (dotted lines). The H$\alpha$ emitters at $z=0.4$ (shown with red circles) are distributed at the top-left corner of this plot, and are well separated from other line emitters at different redshifts (shown with grey scale).

With these procedure, we finally identify 3,085 H$\alpha$ emitters at $z=0.4$ in the DEEP2-3 field. In this paper, we use this unprecedentedly large H$\alpha$ emitter sample selected from a contiguous area on sky to study environmental dependence of the properties of H$\alpha$ galaxies along the cosmic web. We note that the current sample may include AGNs. \citet{hayashi2017} matched the NB emitters in the COSMOS field (selected in the same way as presented here) with deep Chandra X-ray source catalogue available in the COSMOS field (\cite{marchesi2016}), and find that only $\sim$0.1\% of the NB emitters have X-ray counterparts. However, this provides a lower limit of AGN fraction, because not all AGNs have X-ray emission (e.g.\ \cite{garn2010}; \cite{calhau2017}). Broad-line optical AGNs may still have a significant contribution particularly at the luminous end of H$\alpha$ emitters (\cite{sobral2016}). Unfortunately, it is not possible to identify/remove individual AGNs from our H$\alpha$ emitter sample in DEEP2-3 field, and we should keep in mind that some of the environmental trends that we discuss in the following sections may be contributed by AGNs.

\begin{figure*}
 \vspace{0cm}
 \begin{center}
  \includegraphics[width=15cm,angle=0]{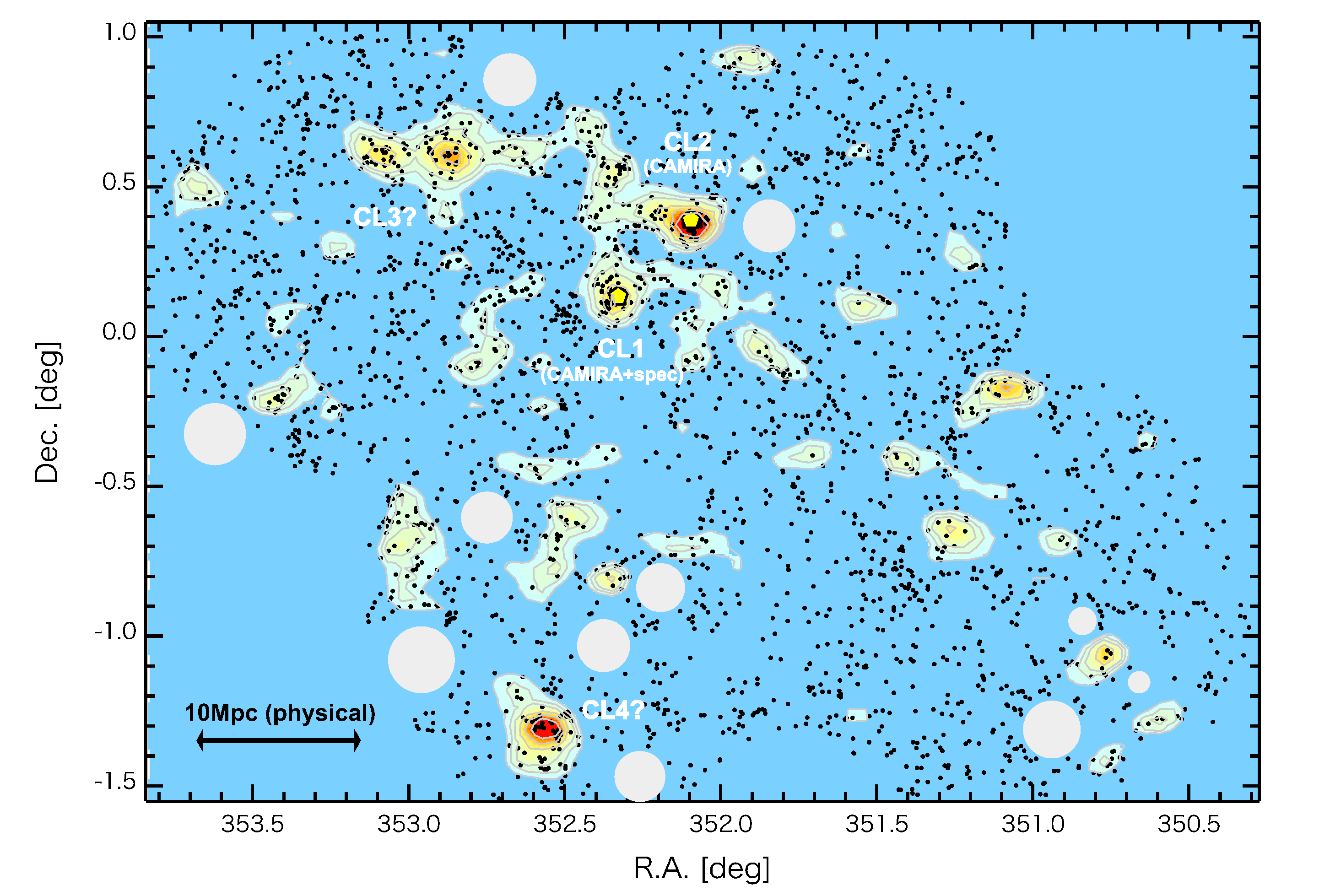} 
 \end{center}
\vspace{-2mm}
\caption{The two-dimensional distribution of H$\alpha$ emitters in the HSC DEEP2-3 field. The small circles indicate the positions of H$\alpha$ emitters. The colour contours indicate 1, 1.5, 2, 3, 4, 5$\sigma$ above the mean density distribution computed with all member galaxies (i.e.\ photo-$z$ selected sample and H$\alpha$ emitters). Here we apply gaussian smoothing for all the data points with $\sigma$$\sim$0.75~Mpc, and coadd the tail of gaussian wing at each position. Two yellow pentagons indicated as CL1 and CL2 show the locations of two CAMIRA clusters at $z\sim 0.4$ identified by the red-sequence finder method by \citet{oguri2017}. Possible strong over-densities are also seen at the north-east and at the south edge of the field (CL3 and CL4). They are also likely in the same structure as CL1/CL2, but the CL3/CL4 are not identified as CAMIRA clusters, and unfortunately there is no galaxies with spectroscopic redshift available around C3/C4 region. Large grey circles show the object masks; we show only large masks with radius of $>$2-arcmin for clarity. }\label{fig:map}
\end{figure*}
\begin{figure}
 \begin{center}
  \includegraphics[width=7.6cm,angle=270]{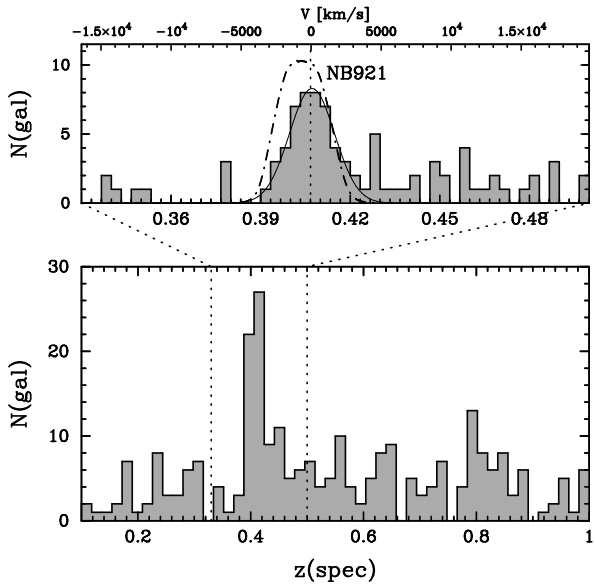} 
 \end{center}
\caption{The redshift distribution of galaxies with spectroscopic redshifts within 1.5-Mpc (or correspondingly 4.6 arcmin) from the centre of CL1. In the top panel, we show the result of gaussian fitting around the redshift spike (solid line). The cluster redshift is determined to be $z=0.4067$, with the velocity dispersion of 1,389~km$\cdot$s$^{-1}$. The dot-dashed line curve shows the NB921 filter transmission curve corresponding to their H$\alpha$ redshift. }\label{fig:zspec_histogram}
\end{figure}

\subsection{Photometric redshift sample}

In addition to the H$\alpha$ emitters selected above, we also make use of the photometric redshift (of non-emitters) to trace large-scale structures at $z=0.4$. Photometric redshifts of HSC sources are published by several authors using independent photo-$z$ codes (Tanaka et al. 2017). In this paper, we use the S15B version of the photo-$z$ from the "DEmP\footnote{Following the recommendation by the  photo-$z$ code developing team, we use the  mode of probability distribution function of $z_{\rm phot}$ for point estimates of photometric redshift.}" code (\cite{hsieh2014}) which seems to provide most reliable redshifts for red galaxies at $z\sim 0.4$, although we verify that our results do not change depending on the choice of photo-$z$ code. 

We note that photometric redshifts estimated from optical 5-band data alone may not be accurate enough, particularly for star-forming galaxies with blue flat SEDs. This trend is seen in our sample as demonstrated in Fig.~\ref{fig:photoz_cut} (right), where we plot the difference between $z_{\rm phot}$ (derived with "DeMP" code) and $z_{\rm spec}$ against their $g-i$ colours for spectroscopically confirmed galaxies at $0.38<z<0.42$ in the DEEP2-3 field available from the DEEP2 redshift survey (\cite{davis2003}; \cite{newman2013}) and the PRIMUS survey (\cite{coil2011}; \cite{cool2013}).

\begin{figure*}
 \begin{center}
 \vspace{-2mm}
  \includegraphics[width=14.0cm,angle=0]{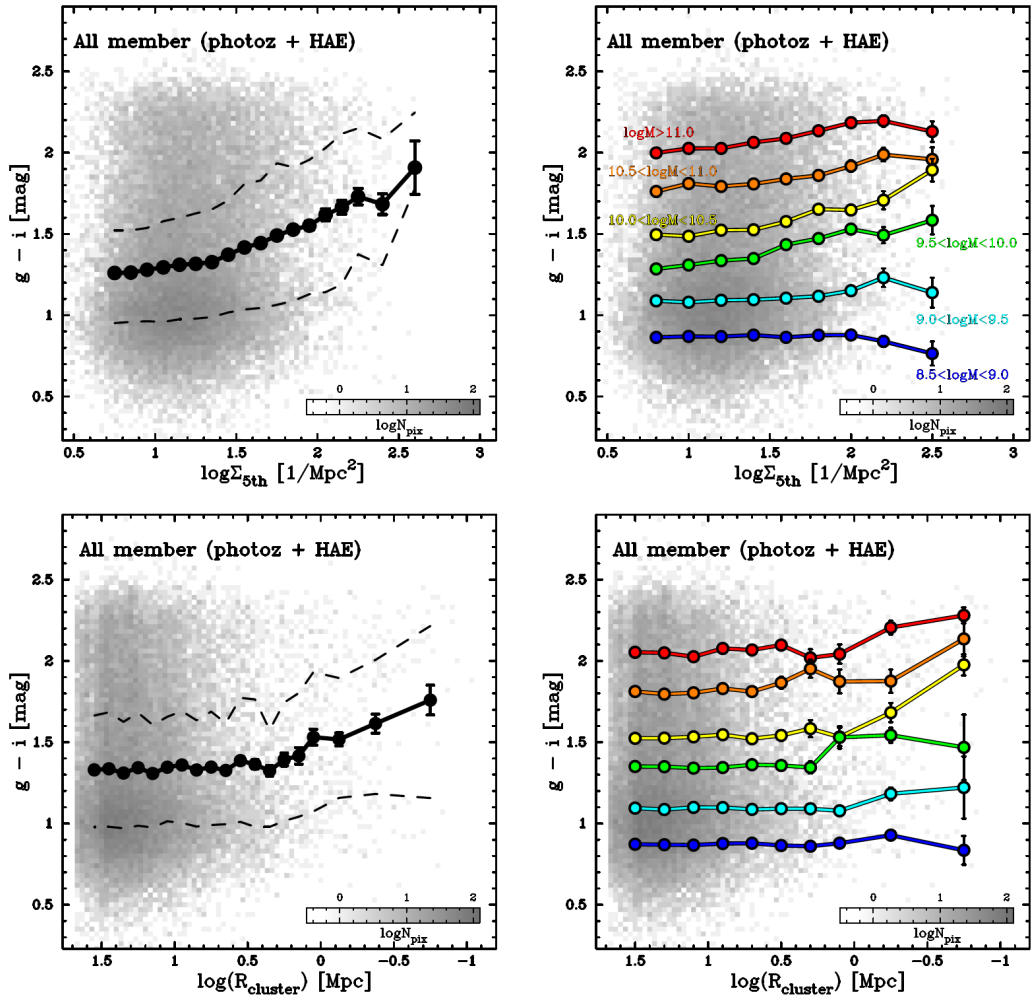} 
 \end{center}
\vspace{-1mm}
\caption{Colours ($g-i$) of all member galaxies plotted against local density $\Sigma_{\rm 5th}$ (top panels) and the distance to the twin clusters (bottom panels). The left panels show the results for all galaxies, with running average ($\pm$standard error computed as $\sigma$/$\sqrt N$ for each subsample). The black dashed lines show the 25- and 75-percentile distribution in each environment bin. Galaxies in higher-density environments (or those with smaller distance to the twin clusters) tend to show redder colours, demonstrating that our environment definition works reasonably well.  The right panels show the results by splitting the sample into six stellar mass bins as indicated in the plot. The trend becomes much milder (or almost disappears) at fixed stellar mass, although there still remains a hint of correlation between colour and environment particularly for massive galaxies. The bin sizes for computing the running average are determined so that each environment bin includes a minimum sample size of $N$$=$5. We follow this requirement in all the similar plots presented in this paper. }\label{fig:color_vs_env_all}
\end{figure*}

In this study, we apply a simple photo-$z$ cut of $z_{\rm phot}$$=$0.40$\pm$0.05 to select "member" galaxies associated to the $z$$=$0.4 structure (as shown with the horizontal dotted lines in Fig.~\ref{fig:photoz_cut}-right). With this photo-$z$ cut, we can select the majority ($\sim$75\%) of real members with $g-i$$>$2.0, while we tend to miss a relatively large fraction ($\sim$60\%) of blue member galaxies with $g-i$$\lesssim$1.5. An important advantage of the NB imaging approach is that we can recover a large fraction of blue star-forming galaxies without broadening the width of photo-$z$ slice. 

Our final sample includes 29,088 galaxies within the photo-$z$ slice of $z_{\rm phot}$$=$0.40$\pm$0.05 (with NB921$<$23.5 mag), and we also include 3,085 H$\alpha$ emitters regardless of their photo-$z$. Because 1,235 ($\sim$40\%) of the H$\alpha$ emitters satisfy the photo-$z$ criteria (i.e. overlapped), our final sample includes 30,938 galaxies in total. The photo-$z$ criteria applied here may still be too wide to trace a single structure, but we need to compromise here; i.e.\ narrower photo-$z$ slice would miss a large fraction of real members, while wider photo-$z$ cut will result in a significant level of contamination from non-member galaxies. We note, however, that our conclusions do not change much even if we apply narrower/broader photo-$z$ cut, because our main goal of this work is to study environmental dependence of H$\alpha$-selected galaxies. In this paper, we use the photo-$z$ selected galaxies to map the large-scale structures and to check their colour--density correlation.

\begin{figure*}
 \begin{center}
 \vspace{-2mm}
 \includegraphics[width=14.0cm,angle=0]{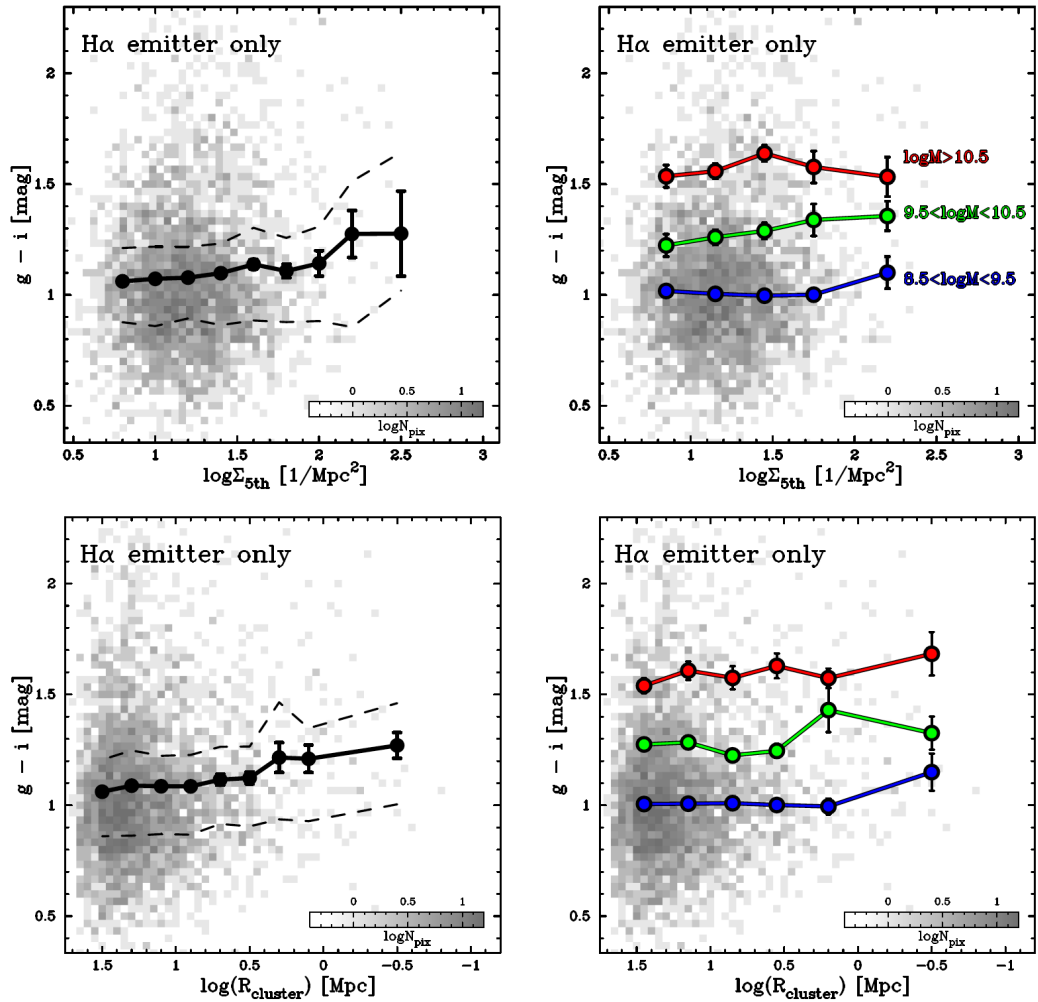} 
 \end{center}
\vspace{-1mm}
\caption{Colours ($g-i$) of H$\alpha$ emitters against their local density (top panels) and the distance to the twin clusters (bottom panels). The left panels show the results for all H$\alpha$ emitters. The black solid line shows the running average ($\pm$standard error computed as $\sigma$/$\sqrt N$), while the black dashed lines show the 25 and 75 percentile distribution. The right panels show the results for H$\alpha$ emitters in three different $M_{\star}$ bins with their running average.}\label{fig:color_vs_env_HAE}
\end{figure*}

\section{Results and Discussion}
\label{sec:result}

\subsection{A super structure hosting twin clusters at z=0.4 in the DEEP2-3 field}
\label{sec:lss}

In Fig.~\ref{fig:map}, we show the locations of H$\alpha$ emitters (dots) on top of the 2-D distribution of all photo-$z$ selected galaxies (colour contours). There are strong density peaks at around CL1$=$(RA1, Dec1)$=$(352.33, $+$0.135) and CL2$=$(RA2, Dec2)$=$(352.096, $+$0.390), both of which are also identified as clusters at $z\sim 0.4$ by \citet{oguri2017} using their red-sequence finding algorithm "{\sc camira}" (Cluster finding Algorithm based on Multi-band Identification of Red-sequence gAlaxies; see \cite{oguri2014}). A strong concentration of H$\alpha$ emitters around the two clusters suggests that the redshift of these two clusters are very close, and they are likely physically associated with each other. There are two similar levels of prominent over-densities at the north-east and at the south edge of the field (shown as CL3 and CL4 in Fig.~\ref{fig:map}), although these structures are not identified as {\sc camira} clusters.

The CL1 region is partly overlapped with the area of DEEP2 redshift survey (\cite{davis2003}; \cite{newman2013}) and PRIMUS survey (\cite{coil2011}; \cite{cool2013}). We show in Fig.~\ref{fig:zspec_histogram} the spec-$z$ distribution of galaxies located within 1.5 Mpc from the CL1 position suggested by \citet{oguri2017}. There is a clear redshift spike at $z=0.4067$, which is consistent with the redshift estimate by \citet{oguri2017}, and is perfectly matched to the H$\alpha$ redshift of the NB921 filter. We find that the velocity dispersion of this cluster is $\sigma$$\sim$1,389~km/s, which can be converted to the cluster mass of $M_{\rm 200}$$\sim$4$\times$$10^{15}$$M_{\odot}$ assuming that the cluster is virialized (\cite{koyama2010}). We caution that the estimated cluster mass may be overestimated, because CL1 is likely an unvirialized system under cluster--cluster merger (with CL2). This is a common problem when deriving the mass of unvirialized clusters using velocity dispersion. An independent measurement with e.g. X-ray observation is needed to determine its cluster mass. 

Although further spectroscopic observation is needed to firmly conclude the physical association between the four candidate clusters reported here, we can at least say that all the density peaks are located within $\pm$2000~km/s, because all of these structures harbour an over-density of H$\alpha$ emitters. We believe that they are organizing a single gigantic structure at $z=0.4$. 

\begin{figure*}
 \begin{center}
 \vspace{-2mm}
  \includegraphics[width=16.0cm,angle=0]{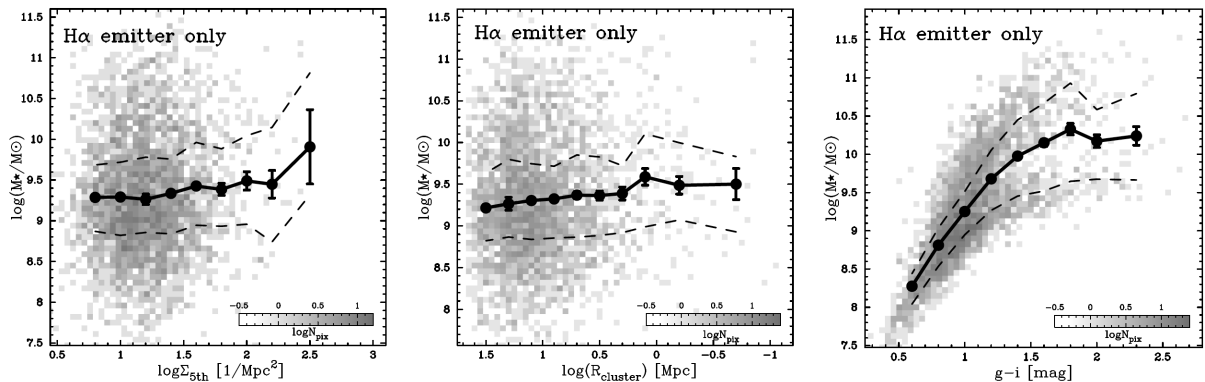} 
 \end{center}
\vspace{-1mm}
\caption{Stellar mass of H$\alpha$ emitters as a function of $\Sigma_{\rm 5th}$ (left), cluster centric radius (middle), and $g-i$ colours (right). The running average (with its standard error), as well as 25 and 75\% distribution, are shown in each panel. There seems to be a marginal trend that $M_{\star}$ increases towards high-density environment, while the significance is very low considering the systematic uncertainties in deriving stellar mass (see Section~3.3). The right-hand panel shows a strong correlation between $M_{\star}$ and colours of H$\alpha$ emitters, suggesting that the possible colour--density and colour--radius correlation for H$\alpha$ emitters reported in Fig.~6 can be at least partly driven by the massive H$\alpha$ emitters in the high-density environment. }\label{fig:Mstar_vs_env}
\end{figure*}

\subsection{Definition of environment}

In this study, we define the environment of each galaxy with two different approaches. The first approach is to use local galaxy number density. By using all the member galaxies (i.e.\ all photo-$z$ selected galaxies and H$\alpha$ emitters), we first compute the distance to its fifth nearest neighbour galaxy for each galaxy ($r_{\rm 5th}$), and define the local density as $\Sigma_{\rm 5th}$$=$5/$\pi r_{\rm 5th}^2$. The second approach is to use the projected distance to the twin clusters discovered in Section~3.1($R_{\rm cl}$). We compute the projected distance from each member galaxy to CL1 and CL2 ($R_1$ and $R_2$), and we define $R_{\rm cl}$$=$min($R_1$, $R_2$). 

We note that there are advantages/disadvantages of each method (see detailed discussion by e.g.\ \cite{muldrew2012}; \cite{darvish2015a}). The $R_{\rm cl}$ approach is simple and easily reproducible, but it is not possible to take into account smaller structures (groups or filaments) along the large-scale structures. For example, the same value of $R_{\rm cl}$ for different galaxies does not necessarily indicate that they belong to similar parent halo. On the other hand, the local density approach is sensitive to all possible structures, while it is also sensitive to the photo-$z$ slice, and will potentially detect over-densities of galaxies at slightly different redshifts as long as we rely on photometric redshifts. We emphasize that the two approaches can complement with each other: the $R_{\rm cl}$ approach can be used to study more global environment, while the $\Sigma_{\rm 5th}$ approach is suited to study local environmental effects.

\subsection{Colour--density and colour--radius correlation with all galaxies at z=0.4}

To check if our environmental definition properly works, we show in Fig.~\ref{fig:color_vs_env_all} the $g-i$ colours of galaxies (which straddle the rest-frame 4000\AA\ break at $z=0.4$) against their local density ($\Sigma_{\rm 5th}$) and the distance to the twin clusters ($R_{\rm cl}$). In the left panels of Fig.~\ref{fig:color_vs_env_all}, we show the results for all galaxies, while the right panels show the results by splitting the sample into six stellar mass bins. Here we derive stellar mass of galaxies using $y$-band  (rest-frame $R$-band) photometry with the following equation:
\begin{equation}
\log (M_*/10^{11}M_{\odot})_{z=0.4} = -0.4(y-19.91) + \Delta\log M_{0.4},
\end{equation}
where the $\Delta\log M_{0.4}$ term indicates the colour dependence of $M/L$ ratio for $z=0.4$ galaxies: 
\begin{equation}
\Delta\log M_{0.4}= 0.045 - 4.60\times \exp[-1.835\times (g-i)].
\end{equation}
These equations are derived using the model galaxies developed by \citet{kodama1999}. This rest-frame $R$-band photometry method is highly reproducible, and reported to work reasonably well at all redshift range out to $z\sim 2$ (with an uncertainty of $\sim$0.3-dex; e.g.\ \cite{koyama2013}).\footnote{We also confirm that the conclusions presented in this paper do not change if we derive stellar mass of H$\alpha$ emitters using the SED-fitting code ``MIZUKI'' (by fixing their redshift at $z=0.4$) developed by one of the authors (\cite{tanaka2015}).}

In Fig.~\ref{fig:color_vs_env_all}, it is shown that galaxies in higher-density environment (or galaxies located closer to the twin clusters) tend to have redder colours, demonstrating that our definitions of environment work reasonably well. It is also demonstrated in Fig.~\ref{fig:color_vs_env_all} (right) that more massive galaxies are always redder in all environments. It seems that there still remains a weak, marginal correlation between colours and environment at fixed stellar mass particularly for galaxies with stellar mass range of 9.5$\lesssim \log(M_{\star}/M_{\odot})$$\lesssim$10.5 (as far as we only focus on the average colours), while low-mass galaxies with $\log(M_{\star}/M_{\odot})$$\lesssim$9.5 are exclusively blue regardless of their environment, although a larger sample would be required to confirm the different behaviour of different stellar mass bins.  

\begin{figure*}
 \begin{center}
 \vspace{-2mm}  
 \includegraphics[width=14.0cm,angle=0]{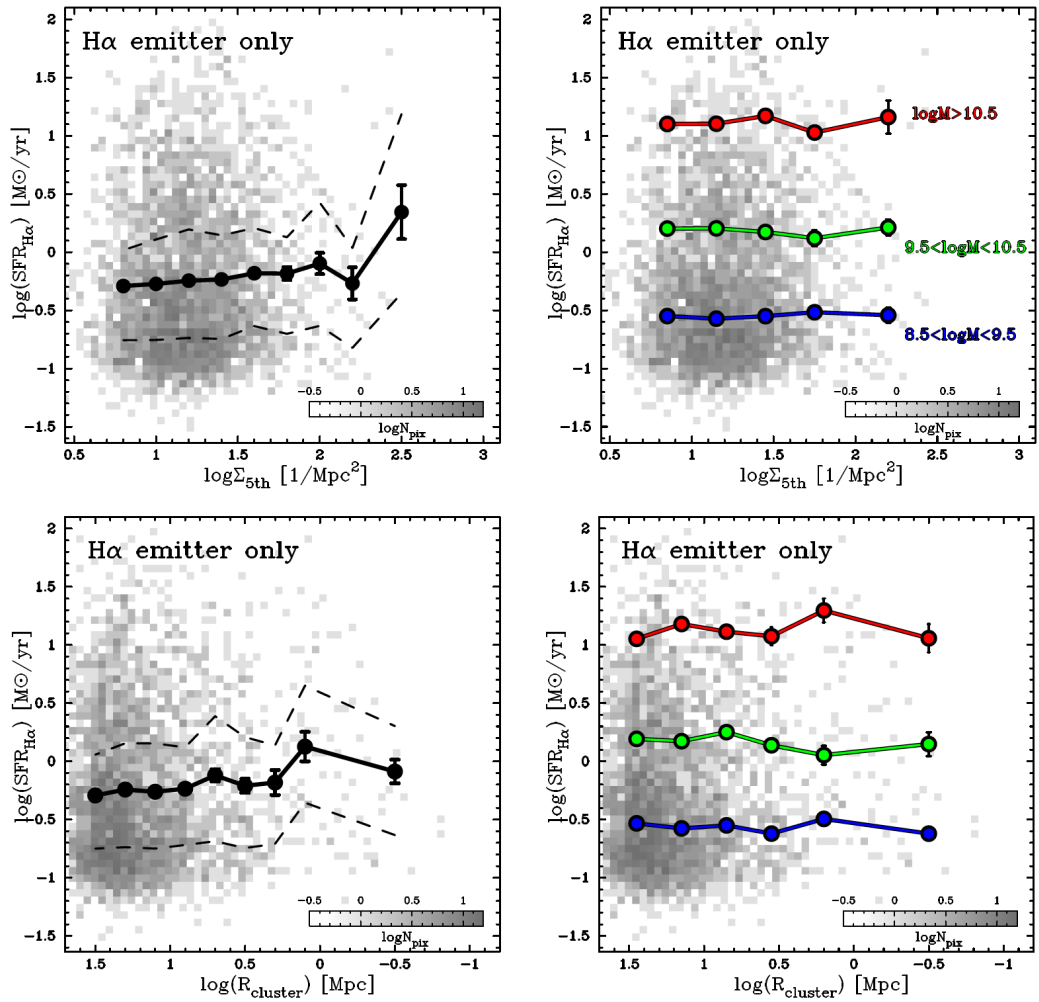} 
 \end{center}
\vspace{-1mm}
\caption{SFR of H$\alpha$ emitters (after dust extinction correction) plotted against their local density (top panels) and the distance to the twin clusters (bottom panels). The left panels show the results for all H$\alpha$ emitters, while the right panels show the result by splitting the sample into three $M_{\star}$ bins as indicated in the panel. The left panels suggest a possible increase of the average SFR towards high-density environments, while the trend becomes much milder (or disappears) at fixed stellar mass (right panels). }\label{fig:sfr_vs_env_HAE}
\end{figure*}
\begin{figure*}
 \begin{center}
 \vspace{-2mm}
 \includegraphics[width=14.0cm,angle=0]{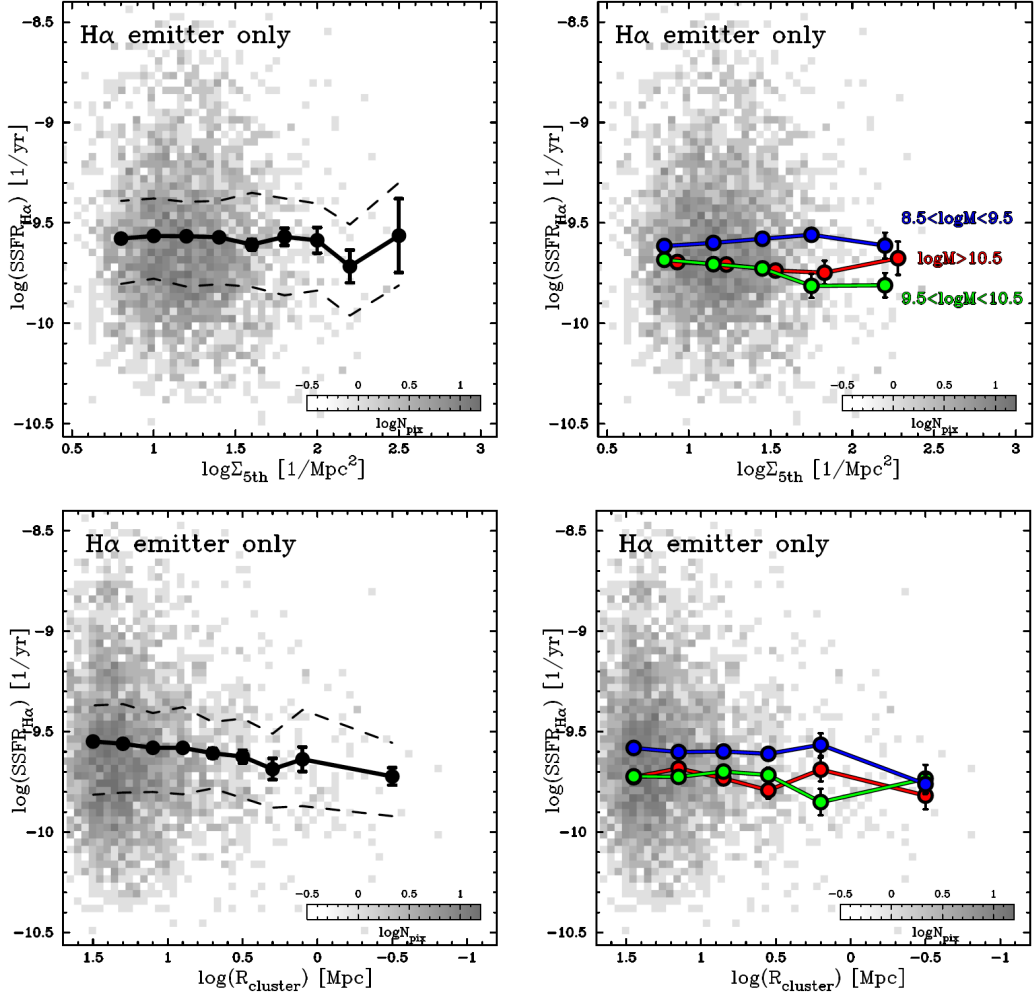} 
 \end{center}
\vspace{-1mm}
\caption{Same plot as Fig.~8, but for the specific SFR of H$\alpha$ emitters. The bottom panels suggest a possible decline in the specific SFR towards high-density, cluster environments (by 0.1--0.2-dex), while the trend becomes much milder (or disappears) at fixed stellar mass (right panels). }\label{fig:ssfr_vs_env_HAE}
\end{figure*}

\subsection{Environmental dependence of colours of H$\alpha$-selected galaxies}
\label{sec:color}

In the reminder of this paper, we study colour, stellar mass, and star forming activity of H$\alpha$-selected galaxies as a function of environment ($\Sigma_{\rm 5th}$ and $R_{\rm cl}$). Fig.~\ref{fig:color_vs_env_HAE} shows $g-i$ colours of H$\alpha$ emitters against their local density (top panels) and cluster centric radius (bottom panels). As we did in Fig.~\ref{fig:color_vs_env_all}, the left panels show the results for all H$\alpha$ emitters, while the right panels show the results for different stellar mass bins. 

Fig.~\ref{fig:color_vs_env_HAE} demonstrates that there might remain a colour--density and colour--radius trend for H$\alpha$-selected galaxies; i.e.\ H$\alpha$ emitters in higher-density environments (or those located closer to the cluster central regions) tend to show redder colours than H$\alpha$ emitters in underdense regions on average, although the trend becomes much milder (or almost disappear) compared with that we reported for all member galaxies in Fig.~\ref{fig:color_vs_env_all}. In particular, the right panels of Fig.~\ref{fig:color_vs_env_HAE} demonstrate that the colour--density and colour--radius relations do not exist any more when we fix their stellar mass. 

The right panels of Fig.~\ref{fig:color_vs_env_HAE} also show that more massive H$\alpha$ emitters have redder colours, irrespective of environment. It is therefore expected that the possible colour--density or colour--radius correlation amongst H$\alpha$ emitters that we marginally see in Fig.~\ref{fig:color_vs_env_HAE} are driven by the stellar mass difference of H$\alpha$ galaxies in different environment. 
To test this possibility, we plot in the left- and middle-panels of Fig.~\ref{fig:Mstar_vs_env} the stellar mass of H$\alpha$ emitters against their environment. Although the statistics is very poor again, as we suspected, H$\alpha$ emitters in higher-density environments tend to be more massive. A question here is whether or not this stellar mass difference can explain the colour--density trend we reported in Fig.~\ref{fig:color_vs_env_HAE}. In the right panel of Fig.~\ref{fig:Mstar_vs_env}, we show the $M_{\star}$ of H$\alpha$ emitters against their $g-i$ colours. It is clear that there is a strong correlation between $M_{\star}$ and colour, and we expect this colour--$M_{\star}$ correlation for H$\alpha$ emitters can be a major driver of the weak colour--density or colour--radius correlation amongst H$\alpha$ emitters reported in Fig.~\ref{fig:color_vs_env_HAE} (see also e.g.\ \cite{hogg2004}; \cite{baldry2006}; \cite{scodeggio2009}; \cite{moresco2010}; \cite{darvish2015a}).

\subsection{H$\alpha$-based star formation activity of galaxies along the huge cosmic web}
\label{sec:sfr}

An important advantage of narrow-band H$\alpha$ imaging studies is that we can measure the SFRs of individual emitters based on the emission line fluxes. We first derive the H$\alpha$+[NII] flux ($F_{\rm H\alpha+[NII]}$), emission-line subtracted continuum flux density ($f_{\rm cont}$), and the rest-frame equivalent width (EW$_{\rm rest}$({H$\alpha$+[NII])) of each H$\alpha$ emitter with the following equations:
\begin{equation}
F_{\rm{H\alpha + [NII]}}=\Delta_{\rm{NB}}\frac{ f_{\rm{NB}} -
 f_{BB} }{1-\Delta_{\rm{NB}}/\Delta_{BB}},
\end{equation}
\begin{equation}
f_{\rm cont} = \frac{f_{BB} - f_{\rm{NB}}(\Delta_{\rm{NB}}/\Delta_{BB})}{1-\Delta_{\rm{NB}}/\Delta_{BB}},
\end{equation}
\begin{equation}
\textrm{EW}_{\rm rest}({\rm{H\alpha + [NII]}}) = (1+z)^{-1} \frac{F_{\rm{H\alpha},
 + [NII]}}{f_{\rm cont}},  
\end{equation}
where $f_{\rm{NB}}$ and $f_{\rm BB}$ denote the flux density of the NB921 and continuum magnitudes derived with Eq. (1) for individual galaxies. The $\Delta_{\rm NB}$($=$135\AA) and $\Delta_{\rm BB}$($=$782\AA) are the band widths of narrow-band (NB921) and broad-band ($z'$) filters. We then multiply 4$\pi d_{L}^2$ to convert the H$\alpha$+[NII] flux to the luminosity ($L_{\rm H\alpha + [NII]}$), where $d_{L}$$=$2.20$\times$10$^3$~Mpc is the luminosity distance at $z=0.4$. 

\begin{figure*}
 \begin{center}
  \vspace{-2mm}
  \includegraphics[width=16.0cm,angle=0]{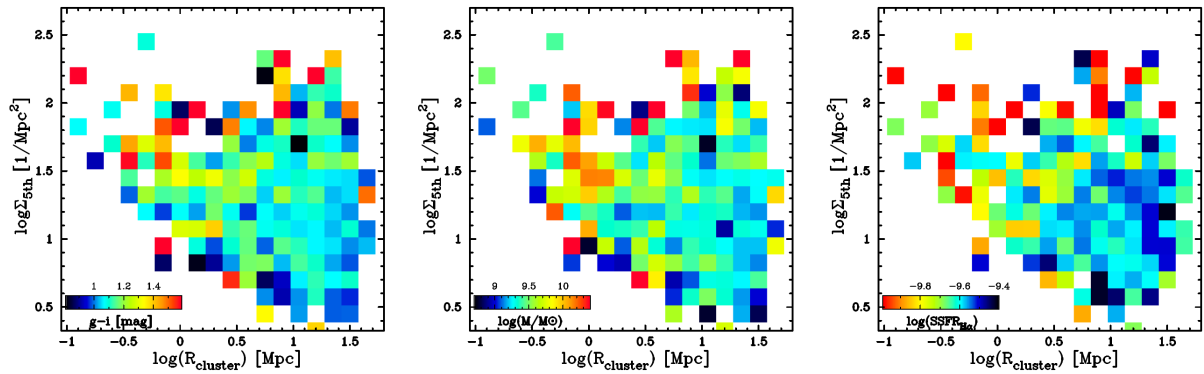}
 \end{center}
\vspace{-1mm}
\caption{The relation between the local galaxy density ($\Sigma_{\rm 5th}$) and cluster centric radius ($R_{\rm cl}$) for H$\alpha$ emitters. The colour coding indicates $g-i$ colour (left), stellar mass (middle), and specific SFR (right). We apply 20$\times$20 gridding and compute average value in each grid. This plot visually demonstrates that both the local galaxy density and the distance to the cluster can affect the physical quantities studied here (colour, mass, sSFR). }\label{fig:Sigma5_vs_Rcl}
\end{figure*}

We need to correct for the [NII] line contribution and dust extinction effect. These steps require complete spectroscopic survey for all the emitters, which is not available. We therefore adopt statistical approaches to use local calibrations to predict [NII] contribution and dust extinction levels. For [NII] correction, we use the correlation between [NII]/H$\alpha$ ratio and EW$_{\rm rest}$(H$\alpha$+[NII]) formulated by \citet{sobral2012}:
\begin{eqnarray}
\log({\rm [NII]/H\alpha}) = -0.924 + 4.802E - 8.892E^2 \nonumber\\ + 6.701E^3  - 2.27E^4 + 0.279E^5,
\end{eqnarray}
where $E$$=$$\log$EW$_{\rm rest}$(H$\alpha$+[NII]). The derived contribution of [NII] lines into total flux ($=$[NII]/(H$\alpha$+[NII])) for the H$\alpha$ emitters are 0.02--0.28 with the median of 0.24. 

For dust extinction correction, we use the recipe to predict H$\alpha$ dust extinction ($A_{\rm H\alpha}$) from stellar mass ($M_{\star}$) and EW$_{\rm rest}$(H$\alpha$) established by \citet{koyama2015}:
\begin{eqnarray}
A_{\rm H\alpha} = a(\log M_{\star})\times \log({\rm EW_{H\alpha}}) + b(\log M_{\star}),
\end{eqnarray}
where $a(\log M_{\star}) = 0.096\times \log M_{\star} -0.717$ and $b(\log M_{\star}) = 0.538\times \log M_{\star} -4.745$ derived by fitting to SDSS star-forming galaxies by taking into account the IMF difference (with typical uncertainties of $\sim$0.3-mag; see \cite{koyama2015} for details).  
The derived $A_{\rm H\alpha}$ ranges 0--2.4 mag with the median $A_{\rm H\alpha}$ of 0.55 mag. After correcting the [NII] contribution and dust extinction, we finally derive the H$\alpha$-based SFR using the \citet{kennicutt1998} calibration assuming the \citet{salpeter1955} IMF, to be consistent with the derivation of stellar mass: 
\begin{equation}
{\rm SFR_{H\alpha}} = 7.9 \times 10^{-42} L_{\rm H\alpha} {\rm [erg/s]}.
\end{equation}
We note that our H$\alpha$ emitter selection criteria described in Section~2 corresponds to the SFR limit of $\sim$0.1~[M$_{\odot}$/yr], and the EW$_{\rm rest}$ limit of $\sim$40\AA.

Fig.~\ref{fig:sfr_vs_env_HAE} shows H$\alpha$-derived SFRs of H$\alpha$ emitters against their local density ($\Sigma_{\rm 5th}$; top panels) and the distance to the twin clusters ($R_{\rm cl}$; bottom panels). Although the trend is not significant, we find a possible increase of average SFR$_{\rm H\alpha}$ of H$\alpha$ galaxies towards high-density cluster environments (i.e.\ the ``reversal'' of SFR--density correlation), likely reflecting the fact that H$\alpha$ emitters in high-density environment are more massive as reported in Fig.~\ref{fig:Mstar_vs_env} and perhaps dustier (\cite{koyama2013}; \cite{sobral2016}). However, we note that this marginal trend disappears when we fix the stellar mass (Fig.~\ref{fig:sfr_vs_env_HAE}-right). We also note that, as we mentioned in Section~2.1, some of the massive, luminous H$\alpha$ emitters in high-density environments which seem to contribute to the different nature of H$\alpha$ galaxies in the highest-density environment could be contributed by AGNs.

We also show in Fig.~\ref{fig:ssfr_vs_env_HAE} the specific SFRs ($=$SFR$_{\rm H\alpha}$/$M_{\star}$) of H$\alpha$ emitters against their environment. The left panels of Fig.~\ref{fig:ssfr_vs_env_HAE} show the results for all H$\alpha$ emitters, while the right panels show the results by splitting the sample into three stellar mass bins. In the left panels of Fig.~\ref{fig:ssfr_vs_env_HAE}, we find no significant environmental difference in the average specific SFRs of H$\alpha$ emitters. There might be a weak, marginal decrease of specific SFR towards high-density cluster environment, but the difference (if any) is only $\sim$0.1--0.2-dex level at maximum. 

As shown in the right panels of Fig.~\ref{fig:ssfr_vs_env_HAE}, it is also important to note that the trend becomes much less significant (or disappears) if we fix their stellar mass. This is another manifestation of the environmental independence of the star-forming main sequence reported by many recent studies (Section~1). We recall that H$\alpha$ emitters in higher-density environment could be more massive (as reported in Fig,~\ref{fig:Mstar_vs_env}), and as shown in the right panels of Fig.~\ref{fig:ssfr_vs_env_HAE}, the specific SFR of massive H$\alpha$ emitters tend to be slightly lower. We therefore suggest that the possible decline in the specific SFR reported in the bottom-left panels of Fig.~\ref{fig:ssfr_vs_env_HAE} is primarily driven by the small excess of red/massive H$\alpha$ emitters in high-density environment as reported in the previous sections.   

\begin{figure*}
 \begin{center}
  \vspace{-2mm}
  \includegraphics[width=14.0cm,angle=0]{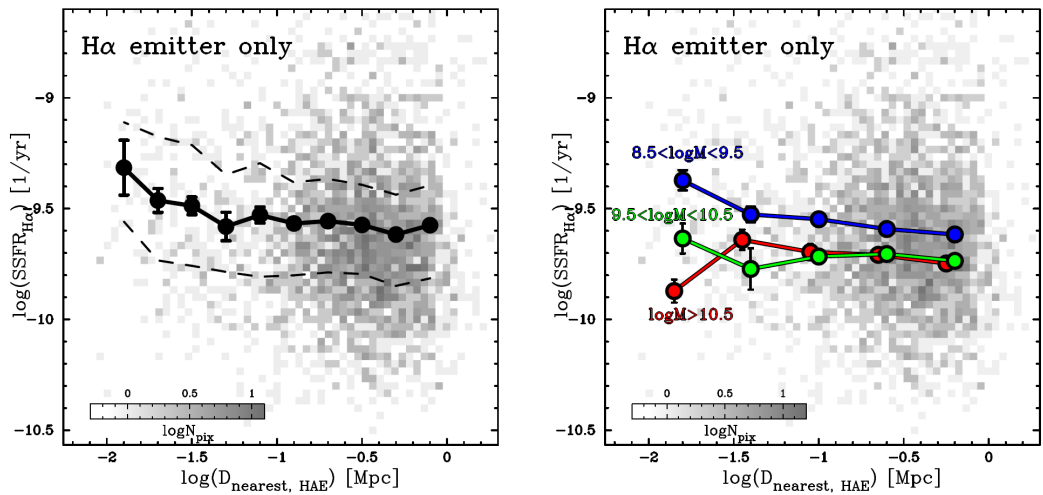} 
 \end{center}
\vspace{-1mm}
\caption{(Left): Specific SFR$_{\rm H\alpha}$ plotted against the distance to the nearest H$\alpha$ emitter of each H$\alpha$ emitter. The meanings of the symbols are the same as the previous figures. (Right): The same plot for three stellar mass bins. }\label{fig:ssfr_vs_D}
\end{figure*}

\subsection{Comparison with other studies on environmental impacts on star-forming galaxies}

In this paper, we suggest a possible environmental dependence of colours of H$\alpha$ emitters in the sense that H$\alpha$ emitters in higher-density environment tend to show redder colours than those in low-density environment (Section~3.4). This result is consistent with our previous work focusing on a single cluster at a similar redshift, where we showed an enhanced red H$\alpha$ emitter fraction in high-density environments (\cite{koyama2011}). We reported that the possible colour--density trend amongst H$\alpha$-selected galaxies seems to be largely driven by the increase of stellar mass of H$\alpha$ emitters towards high-density environment (suggesting a dominant role of stellar mass controlling the environmental dependence of various galaxy properties; \cite{muzzin2012}; \cite{sobral2011}; \cite{darvish2016}; \cite{smethurst2017}). 
We also find that the average SFRs of H$\alpha$ emitters slightly increases towards high-density environment (Section~3.5), which is also consistent with some other studies at $z\sim 1$ (e.g.\ \cite{elbaz2007}; \cite{cooper2008}; \cite{li2011}; \cite{koyama2013}). 

The environmental independence of specific SFR at fixed stellar mass (Section~3.5) is also consistent with many recent studies claiming no or little environmental dependence of the SF main sequence (e.g.\ \cite{balogh2004}; \cite{peng2010}; \cite{wijesinghe2012}; \cite{muzzin2012}; \cite{koyama2013}; \cite{lin2014}; \cite{stroe2015}; \cite{darvish2016}).  
However, we should note that some studies suggest a possible decline in the SF activity of cluster/group SF galaxies both in low-$z$ and high-$z$ universe (\cite{vulcani2010}; \cite{vonderlinden2010}; \cite{haines2013}; \cite{lin2014};  \cite{ziparo2014}; \cite{tran2015}; \cite{allen2016}; \cite{paccagnella2016}; \cite{jian2017}). Although the difference of sSFR between cluster and field galaxies suggested by these studies are not large ($\sim$0.1--0.3-dex levels), it may be important to understand the possible biases associated to each study.  

We expect that there are many possible causes of this discrepancy: e.g. different procedure of SF galaxy selection, different definition of environment, different stellar mass range, or different method of SFR measurement exploited by different authors (see also \cite{darvish2016}; \cite{tran2017}). The NB-based H$\alpha$ approach (like our current work) tends to select galaxies with high EW(H$\alpha$) by nature, and potentially miss low EW(H$\alpha$) galaxies, while an important advantage of H$\alpha$ selected sample is that they do not suffer from contamination due to photo-$z$ error (Section~2). Also, H$\alpha$-based SFRs are advantageous in that it can provide more direct measurements of SFRs compared to SED-based (or colour-based) SFRs, but a potential problem is that there always remain uncertainties regarding the [NII] line contribution and/or dust extinction correction.  

A potentially interesting implication from recent studies (including our current work) is that the effect of environment acts on galaxy colours, mass, or some other spectroscopic properties. For example, dust extinction levels of SF galaxies are higher in higher-density environment (\cite{koyama2013}; \cite{sobral2016}), with potentially higher dust temperatures (\cite{rawle2012}; \cite{matsuki2017}). Some spectroscopic studies also suggest higher gas-phase metallicity in cluster SF galaxies than those in field galaxies (\cite{ellison2009}; \cite{kulas2013}; \cite{shimakawa2015}; \cite{darvish2015b}), although environmental impacts on the chemical abundance or dust properties in star-forming galaxies are still under debate (e.g.\ \cite{patel2011}; \cite{noble2016}; \cite{valentino2016}). Furthermore, \citet{darvish2015b} showed that galaxies in high-density environment show significantly lower electron densities compared with field galaxies. In these ways, our focus is now shifting from the simple comparison between cluster galaxies and field galaxies on the SFR--$M_{\star}$ diagram to unveiling the environmental effects on other important parameters of galaxies. We believe that future spectroscopic survey with the Prime Focus Spectrograph (PFS) on Subaru, or wide-field MIR--FIR survey of this field will allow a more complete census of SF activity in individual galaxies across environment.

\subsection{The role of global and local environment}

We used the local galaxy number density ($\Sigma_{\rm 5th}$) and the distance to the twin clusters ($R_{\rm cl}$) as indicators of environment in this study. For the results presented throughout this section}, it seems that the choice of environmental definition does not strongly affect our results, but it is also interesting to investigate the different role of $\Sigma_{\rm 5th}$ and $R_{\rm cl}$. 
We attempt this in Fig.~\ref{fig:Sigma5_vs_Rcl} by plotting average colour, mass, and sSFR of H$\alpha$ emitters on the $\Sigma_{\rm 5th}$--$R_{\rm cl}$ plane. This plot visually demonstrates that all these galaxy quantities change near the cluster central regions ($R_{\rm cl}$$\lesssim$1~Mpc) and high-density environment ($\Sigma_{\rm 5th}$$\gtrsim$1.8--2.0), suggesting the local and global environment both play an important role (see consistent results by e.g.\ \cite{balogh2004}; \cite{fadda2008}; \cite{alpaslan2016}), although the relative importance of the local/global environment is not fully understood (e.g.\ \cite{kauffmann2004}; \cite{blanton2006}; \cite{cucciati2010}).  

To study the effect of "more local" environment,  we here compute the distance to the nearest neighbour H$\alpha$ emitter ($D_{\rm nearest}$) of individual H$\alpha$ emitters, and study their properties as a function of $D_{\rm nearest}$. This is another advantage of NB imaging survey; this kind of analysis is highly uncertain with photo-$z$ information alone because of severe chance of projection effects. The NB-based H$\alpha$ emitter sample presented in this study are expected to be within a very narrow redshift slice, and we can more reliably use the projected distance to search for a close companion. In the left panel of Fig.~\ref{fig:ssfr_vs_D}, we plot the specific SFR$_{\rm H\alpha}$ of all H$\alpha$ emitters as a function of $D_{\rm nearest}$. Although the trend is weak again, there might be a mild increase of the specific SFR for H$\alpha$ emitters having a close companion emitter within $\sim$30-kpc. This result suggests that SF activity is elevated in interacting systems, while it may be interesting to note that massive H$\alpha$ emitters show different behaviour (see right panel of Fig.~\ref{fig:ssfr_vs_D}). Unfortunately, with the current sample size, it is not possible to go further in detail on the nature of H$\alpha$ emitter pairs (and its relation to their global environment), but we stress that the HSC NB data is ideally suited for this purpose, and it is clearly our important future work after the HSC survey is completed.

\section{Conclusions}
\label{sec:conclusions}

Using the broad-band and narrow-band (NB921) data from the internal release of HSC SSP survey (DR1\_S15B), we present the initial results on the environmental dependence of properties of H$\alpha$-selected galaxies at $z=0.4$ along the cosmic web over $>$5-deg$^2$ in the DEEP2-3 field. Our results are summarized as follows: 

\noindent
(1) By mapping the H$\alpha$ emitters selected by \citet{hayashi2017} and photo-$z$ selected galaxies at $z\sim 0.4$, we identify prominent structures hosting (at least) two clusters at $z=0.4$ in the DEEP2-3 field. These two clusters are independently identified with a red-sequence finder method by \citet{oguri2017}, for one of which we confirmed their physical association by using available $z_{\rm spec}$ information in this paper.
 
\noindent
(2) We define the environment of galaxies using local galaxy number density ($\Sigma_{\rm 5th}$) and the distance to the twin clusters ($R_{\rm cl}$). We confirmed that galaxies in higher-density environments, or galaxies with smaller distance to the twin clusters, tend to show redder colours. 

\noindent
(3) We find a possible trend that H$\alpha$-selected galaxies in higher-density environments or in the cluster central region tend to show redder colours and higher SFR$_{\rm H\alpha}$, likely driven by a mild increase of average stellar mass of H$\alpha$ galaxies towards high-density environment. This result implies that the ``reversal'' of SFR--density correlation reported in the high-$z$ universe can be driven by massive/red H$\alpha$ galaxies in the highest-density environments, although the sample of H$\alpha$ galaxies in the highest-density environments is currently too small. 

\noindent
(4) We find that the average specific SFR$_{\rm H\alpha}$ of H$\alpha$ emitters does not significantly change with environment. Although there might be a marginal decline in the specific SFRs (by 0.1--0.2 dex levels at maximum) towards the highest-density cluster environment, the trend becomes much milder (or disappears) once we fix their stellar mass. This result is consistent with many recent studies claiming environmental independence of SF main sequence. 

\noindent
(5) There is a mild increase of specific SFR of H$\alpha$ emitters for those having close companion emitter within $\lesssim$30~kpc, suggesting an elevated SF activity in galaxies due to galaxy--galaxy interaction processes. 

This paper demonstrated the power of HSC, in particular combined with its narrow-band data, to study galaxy environment in the distant universe.  By incorporating the future data release, we can extend the study towards more complete understanding of the galaxy evolution across a wide environmental range from rich clusters, groups, pairs, to isolated galaxies.


%


\begin{ack}
We thank the referee for their careful reading and constructive comments which improved the paper. 

The Hyper Suprime-Cam (HSC) collaboration includes the astronomical communities of Japan and Taiwan, and Princeton University.  The HSC instrumentation and software were developed by the National Astronomical Observatory of Japan (NAOJ), the Kavli Institute for the Physics and Mathematics of the Universe (Kavli IPMU), the University of Tokyo, the High Energy Accelerator Research Organization (KEK), the Academia Sinica Institute for Astronomy and Astrophysics in Taiwan (ASIAA), and Princeton University.  Funding was contributed by the FIRST program from Japanese Cabinet Office, the Ministry of Education, Culture, Sports, Science and Technology (MEXT), the Japan Society for the Promotion of Science (JSPS),  Japan Science and Technology Agency  (JST),  the Toray Science  Foundation, NAOJ, Kavli IPMU, KEK, ASIAA,  and Princeton University.

This paper makes use of software developed for the Large Synoptic Survey Telescope. We thank the LSST Project for making their code available as free software at  http://dm.lsst.org

The Pan-STARRS1 Surveys (PS1) have been made possible through contributions of the Institute for Astronomy, the University of Hawaii, the Pan-STARRS Project Office, the Max-Planck Society and its participating institutes, the Max Planck Institute for Astronomy, Heidelberg and the Max Planck Institute for Extraterrestrial Physics, Garching, The Johns Hopkins University, Durham University, the University of Edinburgh, Queen’s University Belfast, the Harvard-Smithsonian Center for Astrophysics, the Las Cumbres Observatory Global Telescope Network Incorporated, the National Central University of Taiwan, the Space Telescope Science Institute, the National Aeronautics and Space Administration under Grant No. NNX08AR22G issued through the Planetary Science Division of the NASA Science Mission Directorate, the National Science Foundation under Grant No. AST-1238877, the University of Maryland, and Eotvos Lorand University (ELTE) and the Los Alamos National Laboratory.

This work is based on data collected at the Subaru Telescope and retrieved from the HSC data archive system, which is operated by Subaru Telescope and Astronomy Data Center, National Astronomical Observatory of Japan.

This work was financially supported in part by a Grant-in-Aid for the Scientific Research (Nos.\,26800107, 26707006) by the Japanese Ministry of Education, Culture, Sports and Science.

\end{ack}





\end{document}